\documentclass[12pt, twoside]{article}
\usepackage{a4wide}
\usepackage{latexsym}
\usepackage[centertags]{amsmath}
\usepackage{amsfonts}
\numberwithin{equation}{section}
\begin{document}
\begin{titlepage}
\begin{center}
\hspace{10cm} hep-th/0009083\\
\hspace{10cm} September 2000\\
\vspace*{2cm}

{ \huge Supersymmetric Brane World Scenarios from Off-Shell 
Supergravity}\\
\vspace{1cm}

{\sc Max Zucker}\footnote{e-mail: zucker@th.physik.uni-bonn.de}
\vspace{.2cm}

{\it { Physikalisches Institut \\
Universit\"at Bonn\\
Nussallee 12\\
D-53115 Bonn, Germany}}\\

\vspace{1.2cm}

\vspace{.3cm}

{\large Abstract}
\end{center}
Using $N=2$ off-shell supergravity in five dimensions, we 
supersymmetrize the brane world scenario of Randall and Sundrum. We extend their 
construction to include supersymmetric matter at the fixpoints.

%
%
\end{titlepage}
\section{Introduction}
During the last months, an idea due to Randall and Sundrum 
\cite{Randall:1999ee} has gained a lot of attention. These authors consider five 
dimensional gravity with a cosmological constant on the orbifold 
$S^1/\mathbb{Z}_2$. In addition, there are cosmological constants located at 
the fixpoints of the $\mathbb{Z}_2$. The equations of motion of this 
theory are solved using a warped product ansatz which preserves four 
dimensional Poincar\'e invariance. The exponential dependence of the warp 
factor on the fifth direction then leads to an elegant solution of the 
hierarchy problem. 

Clearly, it would be of great interest to supersymmetrize the 
Randall-Sundrum scenario. To be more precise, one may ask for five dimensional 
gauged supergravity \cite{D'Auria:1981kq} on $S^1/\mathbb{Z}_2$ with 
additional cosmological constants on the branes and supersymmetrizations 
thereof. 

 Some work has already been devoted to the topic of supersymmetric 
brane world scenarios. Let us mention the paper of Falkowski et al. 
\cite{Falkowski:2000er}, which discusses models where the supergravity in the 
bulk is slightly modified; namely the bulk mass term for the gravitino 
contains a step function. Another approach is the work of Altendorfer 
et al. \cite{Altendorfer:2000rr}, who keep the bulk theory as it stands 
and find the terms on the boundaries by requiring the warped product 
metric of Randall-Sundrum being a supersymmetric vacuum. However, their 
method has the disadvantage that it cannot be extended straightforwardly 
to include additional matter on the boundaries. Finally, we should 
mention a more recent preprint, which treats the subject in a more general 
framework \cite{Bergshoeff:2000zn}.

In this work we present an alternative and as we think quite powerful 
technique, suited for the derivation of theories of the Randall-Sundrum 
type. This technique rests mainly on two ingredients. First, Mirabelli 
and Peskin \cite{Mirabelli:1998aj} presented a very elegant method to 
couple theories, which live in the bulk of the orbifold 
$S^1/\mathbb{Z}_2$ to fields located at the fixpoints in a supersymmetric way. However, 
they apply their method to rigidly supersymmetric theories only, which 
is in the light of Randall-Sundrum to restrictive. We stress that the 
method of \cite{Mirabelli:1998aj} rests upon the use of off-shell 
formulations. So the second ingredient is clear: An off-shell formulation of 
gauged supergravity in five dimensions is required in order to 
generalize the idea of Mirabelli and Peskin to local supersymmetry. 

In two recent publications, we have worked out the $N=2$ off-shell 
multiplet calculus in five dimensions. Besides the minimal multiplet and 
the nonlinear multiplet \cite {Zucker:2000ej}, we discussed in 
\cite{Zucker:1999fn} the linear and the super Yang-Mills multiplet. The multiplet 
calculus is completed with the hypermultiplet which may be found in 
\cite{diss} and the tensor multiplet, to be discussed in the following 
section. In a recent publication \cite{Kugo:2000hn} these topics have been 
reconsidered and extended using a different approach.

The purpose of this paper is to use our off-shell calculus in order to 
generalize the idea of Mirabelli and Peskin to local supersymmetry and 
derive in that way lagrangians of the Randall-Sundrum type.

\medskip

The present paper is structured as follows. In section \ref{gos} we 
construct the bulk theory of our orbifold $S^1/\mathbb{Z}_2$, namely 
gauged $N=2$ off-shell supergravity in five dimensions.
In section \ref{sgs} we develop a four dimensional $N=1$ tensor 
calculus at the fixpoints. Using these results we finally study various 
supersymmetric theories on $S^1/\mathbb{Z}_2$ in section \ref{chap4}. Short 
conclusions and an outlook are presented in section \ref{conc}.

\section{Gauged off-shell supergravity\label{gos}}
In this section we construct the gauged supergravity which will be the 
bulk theory for our orbifold construction. 

The fundamental building block is the minimal multiplet. It has been 
discussed in our earlier works, \cite{Zucker:2000ej, Zucker:1999fn}, so 
that we refer to these references for detailed treatments. In order to 
find a consistent supergravity theory, there is a compensator for the 
local $SU(2)_{\cal R}$ required. In principle there are three 
possibilities, leading to three different versions of off-shell supergravity: the 
nonlinear multiplet \cite{Zucker:2000ej} (version I), the 
hypermultiplet \cite{diss} (version II) and the tensor multiplet (version III) which 
is discussed in this work and will be used for the orbifolding 
procedure. These are all (minimal) off-shell versions of $N=2$ supergravity 
which exist in five dimensions. These results are easily extended to 
gauged supergravity. 

A detailed presentation of the three versions of off-shell 
supergravity, including explicit expressions for the lagrangians, may be found in 
\cite{diss}. An analysis of the behavior of the gauged theories under 
$\mathbb{Z}_2$ orbifolding is included in that work. 

\subsection{The tensor multiplet}\label{tenm}
We start by reminding the reader of the linear multiplet 
\cite{Zucker:1999fn}. It contains a Lorentz scalar isotriplet $\vec{Y}$, a spinor 
$\rho$, a scalar $N$ and a vector $W^A$. The vector $W^A$ is constrained 
(for detailed formulas see \cite{Zucker:1999fn} and for the centrally 
charged multiplet \cite{diss})\footnote{We use the same conventions as in 
\cite{Zucker:2000ej}, with the exception that for five dimensional 
indices we use capital letters and lower case letters for four dimensional 
indices. Letters from the middle of the alphabet are curved, letters 
from the beginning of the alphabet are flat.} 
\begin{equation}
\partial_A W^A+\ldots=0.\label{cons}
\end{equation}
As in four dimensional conformal supergravity \cite{deWit:1983na}, one 
may solve this constraint explicitly for vanishing gauge group and 
vanishing central charge. We can achieve that by  the introduction of a 
3-form tensor potential through
\begin{equation}
W^A = \frac{1}{12}e_M^A\varepsilon^{MNPQR}\widehat{\cal D}_N 
B_{PQR}\label{dual}.
\end{equation}
This field forms together with the remaining fields of the linear 
multiplet the tensor multiplet. Schematically, the tensor multiplet is then 
given by
\[
(~\vec{Y},~\rho,~ B_{MNP},~ N).
\]
The introduction of the field $B_{MNP}$ implies a new symmetry, namely 
tensor gauge transformations
\begin{equation}
\delta_\Lambda(\lambda)B_{MNP}=3\partial_{[M}\lambda_{NP]}.\label{tenseich} 
\end{equation}
From the constraint (\ref{cons}) it is possible to deduce the 
supersymmetry transformation law of the tensor field. One finds
\[
\delta B_{MNP} = 
-i\bar{\varepsilon}\gamma_{MNP}\rho-3\bar{\varepsilon}\vec{\tau}\gamma_{[NP}\psi_{M]}\vec{Y}.
\]
It then follows that the supercovariant derivative in (\ref{dual}) is 
given by 
\[
\widehat{\cal D}_{[M} B_{NPQ]} = \partial_{[M} 
B_{NPQ]}+i\bar{\rho}\gamma_{[NPQ}\psi_{M]}-\frac{3}{2}\bar{\psi_{[N}}\vec{\tau}\gamma_{PQ}\psi_{M]}\vec{Y}.
\]
The tensor gauge transformations (\ref{tenseich}) appear in the 
supersymmetry algebra. The additional terms are
\[
[\delta_Q(\eta),\delta_Q(\varepsilon)]=\ldots + 
\delta_\Lambda(i\bar{\varepsilon}\gamma^P\eta 
B_{MNP}-\bar{\varepsilon}\vec{\tau}\gamma_{MN}\eta\vec{Y})
\]
and the dots denote field dependent transformations which may be found 
in \cite{Zucker:2000ej}.

\subsection{The action for the tensor multiplet}\label{tenact}
An action for the tensor multiplet is found as follows 
\cite{deWit:1983na, Bergshoeff:1986mz}: one starts with the coupling of a Maxwell 
multiplet to the tensor multiplet. The tensor multiplet is inert under the 
gauge group. One easily finds that
\begin{equation}
\begin{split}
{\cal L} & =  2 
\vec{X}\vec{Y}-2i\bar{\Omega}\rho-2MN-\frac{1}{12}\varepsilon^{ABCDE}G_{AB}B_{CDE}-16\vec{Y}\vec{t}M\\ & 
-i\bar{\rho}\gamma^M\psi_M M 
+\frac{1}{2}\bar{\psi}_A\vec{\tau}\gamma^{AB}\psi_B\vec{Y}M-\bar{\Omega}\vec{\tau}\gamma^M\psi_M\vec{Y}\label{tensoraction}
\end{split}
\end{equation}
is invariant. Next, one forms a vector multiplet out of the fields of 
the tensor multiplet. Here one faces the first problem. The canonical 
embedding should start with $M\sim N+\ldots$ as it does in the case of 
rigid supersymmetry. However, this cannot be generalized 
straightforwardly to the local case \cite{deWit:1983na}. Instead, one has to move on to 
the improved tensor multiplet \cite{deWit:1983na, deWit:1982fh}.

Since the correspondence is rather complicated, we give only the lowest 
component $M$ of this vector multiplet (the full formulas may be found 
in \cite{diss}):
\begin{equation}
M = 
Y^{-1}N+\frac{1}{4}Y^{-3}\bar{\rho}\vec{\tau}\rho\vec{Y}+6Y^{-1}\vec{Y}\vec{t}.\label{emb1}
\end{equation}
Here we have defined 
\[
Y=(\vec{Y}\vec{Y})^{1/2}
\]
and let us mention that similar cohomology problems as in four 
dimensions \cite{deWit:1983na} appear for the gauge field strength but are 
unimportant for our purposes. Acting repeatedly with the supersymmetry 
transformations on expression (\ref{emb1}) gives the complete embedding.

Using this embedding of the tensor multiplet in the vector multiplet in 
the action formula (\ref{tensoraction}) gives the desired action for 
the tensor multiplet:
\begin{equation}
\begin{split}
{\cal L}_{tensor} = & 
-\frac{1}{4}YR(\widehat{\omega})^{AB}{}_{AB}+4YC-\frac{1}{6}Y\widehat{F}_{AB}\widehat{F}^{AB}+Yv_{AB}v^{AB}+20 
Y\vec{t}^2  -  Y^{-1}N^2\\
&  - 36Y^{-1}(\vec{t}\vec{Y})^2 -\frac{1}{4}Y^{-1}\widehat{\cal 
D}_A\vec{Y}\widehat{\cal D}^A\vec{Y}- 
\frac{1}{12}Y^{-1}\varepsilon^{MNPQR}\vec{Y}\vec{V}_M\partial_N B_{PQR}\\ 
& + Y^{-1}W_A W^A- 4Y^{-1}\bar{\lambda}\vec{\tau}\rho\vec{Y} 
-  
2iY\bar{\psi}_A\gamma^A\lambda-\frac{i}{4\sqrt{3}}Y^{-1}\bar{\rho}\gamma^{AB}\rho\widehat{F}_{AB}\\
& - \frac{i}{2}Y^{-1}\bar{\rho}\gamma^A {\cal 
D}_A\rho-3Y^{-3}\bar{\rho}\vec{\tau}\rho\vec{Y}\tilde{t}\tilde{Y} - 
\frac{1}{2}Y^{-3}\bar{\rho}\vec{\tau}\rho\vec{Y}N-\frac{i}{4}Y^{-1}\bar{\rho}\gamma^{AB}\rho 
v_{AB}\\
& +\frac{1}{4}Y^{-3}\bar{\rho}\vec{\tau}\gamma^A\rho(\vec{Y}\times 
\widehat{\cal D}_A\vec{Y}) 
+\frac{1}{2}Y^{-1}\vec{Y}\bar{\psi}_A\vec{\tau}\gamma^{AB}{\cal D}_B\rho+ iY^{-1}\bar{\rho}\psi_A W^A\\
&  + \frac{1}{2}Y^{-3}\bar{\rho}\vec{\tau}\gamma^A\rho \vec{Y}W_A - 
\frac{1}{2}Y^{-1}\bar{\rho}\vec{\tau}\gamma^{MN}{\cal D}_M\psi_N 
\vec{Y}-\frac{i}{2}Y\bar{\psi}_P\gamma^{PMN}{\cal D}_M\psi_N\\ 
& -12 Y^{-1}N\vec{t}\vec{Y}-Y\bar{\psi}_A\vec{\tau}\gamma^{AB}\psi_B 
\vec{t}-\frac{i}{2}Y\bar{\psi}_A\psi_B 
v^{AB}+2Y^{-1}\bar{\psi}_A\vec{\tau}\gamma^A\rho(\vec{t}\times\vec{Y})\\
& + 
\frac{1}{24}Y^{-3}\varepsilon^{MNPQR}\vec{Y}(\partial_M\vec{Y}\times\partial_N\vec{Y})B_{PQR}
-  
\frac{i}{4\sqrt{3}}Y\bar{\psi}_A\gamma^{ABCD}\psi_B\widehat{F}_{CD}\\
& + Y^{-1}\bar{\rho}\vec{\tau}\gamma_B\psi_A v^{AB}\vec{Y} 
+2iY^{-1}\bar{\rho}\gamma^A\psi_A\vec{t}\vec{Y}- 
\frac{1}{2\sqrt{3}}Y^{-1}\bar{\rho}\vec{\tau}\gamma^{ABC}\psi_A\widehat{F}_{BC}\vec{Y}\\
&  - 
\frac{1}{4}Y^{-1}\bar{\psi}_A\vec{\tau}\gamma^{ABC}\psi_B(\vec{Y}\times\widehat{\cal D}_C\vec{Y}) - 
\frac{1}{2}Y^{-3}\bar{\psi}_A\vec{\tau}\gamma^{AB}\rho\vec{Y}\tilde{Y}\widehat{\cal 
D}_B\tilde{Y}+\frac{1}{8}Y^{-3}(\bar{\rho}\vec{\tau}\rho)^2\\
& 
-\frac{i}{2}Y^{-3}\bar{\psi}_A\gamma^A\rho\bar{\rho}\vec{\tau}\rho\vec{Y}+\frac{1}{2}Y^{-1}\bar{\rho}\psi_B\bar{\rho}\gamma^{AB}\psi_A-\frac{i}{4}Y^{-1}\bar{\rho}\psi_B\bar{\psi}_A\vec{\tau}\gamma^{ABC}\psi_C\vec{Y}\\
& -\frac{3}{8}Y^{-5}(\bar{\rho}\vec{\tau}\rho\vec{Y})^2 + 
\frac{1}{8}Y^{-3}\bar{\psi}_A\vec{\tau}\gamma^{AB}\psi_B\bar{\rho}\tilde{\tau}\rho\vec{Y}\tilde{Y} 
+ 
\frac{i}{4}Y^{-3}\bar{\psi}_A\vec{\tau}\tilde{\tau}\gamma^A\rho\bar{\rho}\tilde{\tau}\rho\vec{Y}\\
&  
-\frac{i}{4}Y^{-3}\bar{\rho}\vec{\tau}\gamma_B\rho\bar{\rho}\gamma^{AB}\psi_A\vec{Y} - 
\frac{1}{8}Y^{-3}\bar{\rho}\vec{\tau}\gamma_B\rho\bar{\psi}_A\tilde{\tau}\gamma^{ABC}\psi_C\vec{Y}\tilde{Y}.\label{tensact}
\end{split}
\end{equation}
\subsection{An action for gauged supergravity} \label{sugact}
Using the results of the preceding section we now construct an action 
for gauged off-shell supergravity following \cite{deWit:1983na}. As 
compensator for the $SU(2)_{\cal R}$ ${\cal R}$-symmetry we use the 
$\vec{Y}$-field. However, before dealing with the gauge fixing, we have to 
consider an old problem: The appearance of a term linear in $C$ in eq. 
(\ref{tensact}) leads to inconsistent equations of motion. We can solve 
this problem in an elegant way, using the lagrangian ${\cal L}_{min}$ for 
the minimal multiplet, eq. (2.10) in \cite{Zucker:2000ej}. The 
lagrangian ${\cal L}={\cal L}_{tensor}+{\cal L}_{min}$ leads to consistent 
equations of motion. It is a well defined lagrangian for off-shell 
supergravity.

The gauged variant of this theory is also easily found: Using the 
lagrangian ${\cal L}_{lin}$ for the linear multiplet given in 
\cite{Zucker:1999fn}, eq. (3.2), the desired lagrangian is 
\begin{equation}
{\cal L}_{gauged}={\cal L}_{tensor}+{\cal L}_{min}-\frac{\sqrt{3}}{4} 
g'{\cal L}_{lin}.\label{orbi1}
\end{equation}
Here we have introduced the cosmological constant $g'$. Of course, 
wherever $W^A$ appears it has to be understood as supercovariant field 
strength for $B_{MNP}$, defined by (\ref{dual}).

Before we proceed, we should discuss the gauge fixing. In our 
formulation of off-shell supergravity, we have an $SU(2)_{\cal R}$ ${\cal 
R}$-symmetry which is gauged by an auxiliary field $\vec{V}_M$. This symmetry 
has to be fixed. We do that by setting \cite{deWit:1983na}
\begin{equation}
\vec{Y}=e^u(0,1,0)^T\label{eichfixx}
\end{equation}
where we have introduced a new scalar $u$. This breaks the original 
$SU(2)_{\cal R}$ but leaves a residual $SO(2)$, gauged by $V_M^2$, intact. 
In addition, there is still the $U(1)$ under which only the graviphoton 
$A_M$ transforms at this stage. The equations of motion for the 
auxiliary fields imply $V_M^2=2g'A_M$ so that after elimination of the 
auxiliary fields, we end up with the lagrangian of gauged supergravity given 
in \cite{Zucker:1999fn}. 

Unfortunately, the lagrangian (\ref{orbi1}) after the gauge fixing 
(\ref{eichfixx}) is quite complicated. Nevertheless it is the starting 
point for our orbifold construction so that we should present explicit 
formulas:
\begin{equation}
\begin{split}
{\cal L}_{gauged} = & 
e^u(-\frac{1}{4}R(\widehat{\omega})_{AB}{}^{AB}+4C-\frac{1}{6}\widehat{F}_{AB}\widehat{F}^{AB}+v_{AB}v^{AB}+20 
\vec{t}\vec{t}-36{(t^2)}^2\\
& -\frac{1}{4}\partial^Au\partial_Au - \frac{1}{4}V_A^1V^{A1} - 
\frac{1}{4}V_A^3V^{A3}+8\sqrt{3}g't^2-\frac{i}{2}\bar{\psi}_P\gamma^{PMN}{\cal 
D}_M\psi_N\\
&-2i\bar{\psi}_A\gamma^A\lambda-\frac{\sqrt{3}g'}{4}\bar{\psi}_A\tau^2\gamma^{AB}\psi_B 
- \frac{i}{2}\bar{\psi}_A\psi_Bv^{AB}) -12Nt^2+\sqrt{3}g'N\\
& -\frac{1}{\sqrt{3}}F_{AB}v^{AB} 
-\frac{1}{6\sqrt{3}}\varepsilon^{ABCDE} A_AF_{BC}F_{DE}- 
4\bar{\lambda}\tau^2\rho-2i\bar{\lambda}\gamma^A\psi_A\\
& + 
\frac{1}{2}\bar{\rho}\tau^2\psi_A\partial^Au-\frac{1}{2}\bar{\rho}\tau^1\psi^MV_M^3+\frac{1}{2}\bar{\rho}\tau^3\psi^MV_M^1+\frac{1}{2}\bar{\psi}_A\tau^2\gamma^{AB}{\cal 
D}_B\rho\\
& + 2i\bar{\rho}\gamma^A\psi_A t^2 - 2\bar{\psi}_A\tau^1\gamma^A\rho 
t^3 + 2\bar{\psi}_A\tau^3\gamma^A\rho 
t^1-\frac{1}{2}\bar{\psi}_A\tau^2\gamma^{AB}\rho \partial_B u\\
& - \frac{1}{12}\varepsilon^{MNPQR}(V_M^2-2g'A_M)\partial_NB_{PQR}-32 
\vec{t}\vec{t}-\frac{\sqrt{3}ig'}{2}\bar{\psi}_A\gamma^A\rho \\
& - \frac{1}{2}\bar{\rho}\tau^2\gamma^{MN}{\cal D}_M\psi_N + 
\bar{\rho}\tau^2\gamma_B\psi_Av^{AB} - 
\frac{1}{2\sqrt{3}}\bar{\rho}\tau^2\gamma^{ABC}\psi_A\widehat{F}_{BC}\\
& - 
\frac{i}{4\sqrt{3}}\bar{\psi}_A\gamma^{ABCD}\psi_B(e^u\widehat{F}_{CD} + \frac{1}{2}F_{CD}) + 
(1-e^u)\bar{\psi}_A\vec{\tau}\gamma^{AB}\psi_B\vec{t}\\
& +e^{-u}\big( - 
\frac{i}{4}\bar{\rho}\gamma^{AB}\rho(v_{AB}+\frac{1}{\sqrt{3}}\widehat{F}_{AB}) - 3\bar{\rho}\tau^2\rho t^2+W_AW^A-N^2\\
& -\frac{i}{2}\bar{\rho}\gamma^A{\cal D}_A\rho+i\bar{\rho}\psi_A 
W^A\big)-4C -\frac{1}{2}e^{-2u}(\bar{\rho}\tau^2\rho N 
-\bar{\rho}\tau^2\gamma^A\rho W_A) +{\cal L}_{4F}.
\end{split}
\end{equation}
${\cal L}_{4F}$ contains four fermion terms which play no r\^ole for 
us. They may be found in \cite{diss}. The covariant derivatives which 
appear here are covariant w.r.t. local Lorentz and local $SO(2)$ 
transformations, e.g.
\[
{\cal 
D}_M\rho=\partial_M\rho+\frac{1}{4}\widehat{\omega}_{MAB}\gamma^{AB}\rho-\frac{i}{2}\tau^2\rho V_M^2.
\]
Let us stress, that we work in what follows till the end with the 
gauged $SU(2)_{\cal R}$. The very last step is to impose the condition 
(\ref{eichfixx}).
\section{Supergravity on $S^1/\mathbb{Z}_2$}\label{sgs}
We now consider gauged supergravity on $S^1/\mathbb{Z}_2$. The 
$\mathbb{Z}_2$ acts on the fifth coordinate, $x^5\to -x^5$. A generic bosonic 
field transforms like
\[
\varphi(x^m,x^5)\to{\cal P}\varphi(x^m,-x^5),\qquad \mbox{with}\qquad 
{\cal P}^2=1.
\]
Fermionic fields transform like
\begin{equation}
\psi(x^5)\to {\cal P}i\tau^3\gamma^{\dot{5}}\psi(-x^5).\label{proj}
\end{equation}
An extended discussion of $\mathbb{Z}_2$ assignments to symplectic 
Majorana spinors can be found in \cite{Bergshoeff:2000zn}. We use the 
notation that a dot on an index $5$ denotes a Lorentz index.
\subsection{The minimal multiplet}\label{sugzs}
Starting from ${\cal P}(e_m^a)=+1$, the parity assignments of all 
fields belonging to the minimal multiplet are dictated by supersymmetry. We 
have collected these assignments in table \ref{table3022}.
\begin{table}
\begin{center}
\begin{tabular}{|c|c|c|c|c|c|c|c|c|c|c|}
\hline\hline
Field & $e_m^a$ & $e_m^{\dot{5}}$ & $e_5^a$ & $e_5^{\dot{5}}$ & 
$\psi_m$ & $\psi_5$ & $A_m$ & $A_5$ & $V_m^1$&  $V_m^2$\\
\hline
Parity $\cal P$&$+1$&$-1$&$-1$&$+1$&$+1$&$-1$&$-1$&$+1$&$-1$&$-1$\\
\hline
\end{tabular}
\begin{tabular}{|c|c|c|c|c|c|c|c|c|c|c|c|}
\hline
Field&  $V_m^3$& $V_5^1$ & $V_5^2$ & $V_5^3$ &$v_{ab}$ & $v_{a\dot{5}}$ 
& $\lambda$ & $C$ & $t^1$& $t^2$ & $t^3$ \\
\hline
Parity $\cal 
P$&$+1$&$+1$&$+1$&$-1$&$-1$&$+1$&$+1$&$+1$&$+1$&$+1$&$-1$\\
\hline\hline
\end{tabular}
\end{center}
\caption{Parities of the minimal multiplet.}\label{table3022}
\end{table}
Note that the orbifold condition breaks the $SU(2)_{\cal R}$ at the 
fixpoints to a residual $U(1)$.

Let us define
\begin{subequations}
\begin{eqnarray}
b_a & = &  v_{a\dot{5}}\\
a_m & = & -\frac{1}{2}( V_m^3 - \frac{2}{\sqrt{3}} 
\widehat{F}_{m5}e^5_{\dot{5}}+4e_m^av_{a\dot{5}})\\
S & = & C-\frac{1}{2}e_{\dot{5}}^5(\partial_5 
t^3-\bar{\lambda}\tau^3\psi_5+V_5^1 t^2-V_5^2t^1)
\end{eqnarray}\label{wichtig1}
\end{subequations}
at the fixpoints. The important observation is then that the fields 
\[
\big( e_m^a,~ \psi_m,~ b_a,~ a_m,~ \lambda,~ S, ~t^1,~t^2\big),
\]
taken at the boundaries, form a non-minimal $N=1$ supergravity 
multiplet in four dimensions with $(16+16)$ components. This multiplet, 
sometimes called the intermediate multiplet has been studied in detail by 
Sohnius and West \cite{Sohnius:1983xs}. These authors also constructed the 
corresponding multiplet calculus.

The $U(1)$ which survives the orbifold projection is to be identified 
with the chiral $U(1)$ gauged by the auxiliary field $a_m$. The fields 
transform as follows under this symmetry (with parameter $\alpha$):
\begin{gather*}
\delta\psi_m=\gamma^{\dot{5}}\psi_m\alpha,\qquad 
\delta\lambda=\gamma^{\dot{5}}\lambda\alpha,\\
\delta a_m=\partial_m\alpha,\qquad \delta t^1=-2t^2\alpha,\qquad \delta 
t^2=2t^1\alpha.
\end{gather*}

The supersymmetry transformation laws on the boundary are in our 
notations
\begin{equation}
\begin{split}
\delta e_m^a & =  -i\bar{\varepsilon}\gamma^a\psi_m\\
\delta \psi_m & =  {\cal D}_m\varepsilon -3\gamma^{\dot{5}}\varepsilon 
b_m 
+\gamma_m(\gamma^a\gamma^{\dot{5}}b_a+2i\tau^1t^1+2i\tau^2t^2)\varepsilon\\
\delta b_a & =  
-\frac{i}{4}\bar{\varepsilon}\gamma_a{}^{bc}\gamma^{\dot{5}}\widehat{\cal 
R}_{bc}-2i\bar{\varepsilon}\gamma_a\gamma^{\dot{5}}\lambda\\
\delta a_m & =  
-\frac{i}{2}\bar{\varepsilon}\gamma^{\dot{5}}\gamma_m\lambda'\\
\delta \lambda &  =  -\frac{1}{4}\gamma^{\dot{5}}\varepsilon 
\widehat{{\cal D}}_ab^a+\varepsilon S 
-\frac{i}{2}\tau^1\gamma^a\varepsilon\widehat{{\cal D}}_a t^1-\frac{i}{2}\tau^2\gamma^a\varepsilon\widehat{{\cal 
D}}_a t^2\\
\delta S & =  -\frac{i}{2}\bar{\varepsilon}(\gamma^m\widehat{{\cal 
D}}_m +\gamma^a\gamma^{\dot{5}} b_a)\lambda 
-\bar{\varepsilon}(\tau^1t^1+\tau^2t^2)(\frac{1}{2}\gamma^{ab}\widehat{\cal R}_{ab}+8 \lambda)\\
\delta t^1 & =  \bar{\varepsilon}\tau^1\lambda\\
\delta t^2 & =  \bar{\varepsilon}\tau^2\lambda.
\end{split}\label{mino}
\end{equation}
Note that fields like $\psi_5$ and $A_5$ have dropped out. The 
covariant derivatives which appear here are covariant w.r.t. four dimensional 
local Lorentz transformations and chiral $U(1)$ transformations. 
Further, we have introduced the useful definition \cite{Sohnius:1983xs}
\[
\lambda'=12\lambda+\gamma^{ab}\widehat{\cal R}_{ab}.
\] 
$\lambda'$ transforms quite simple under supersymmetry:
\[
\delta \lambda'=12\varepsilon S-\frac{1}{4}\varepsilon 
\widehat{R}(\widehat{\omega})^{ab}{}_{ab}+48 \varepsilon ((t^1)^2+(t^2)^2)-6\varepsilon 
b_ab^a-\frac{1}{2}\gamma^{ab}\gamma^{\dot{5}}\varepsilon\widehat{f}_{ab}
\]
and $\widehat{f}_{mn}$ is the supercovariant field strength of $a_m$. 
We remind the reader that in these equations, the fermions $\psi_m$ and 
$\lambda$ satisfy the chirality constraints 
\[
\lambda=i\tau^3\gamma^{\dot{5}}\lambda\qquad\mbox{and}\qquad  
\psi_m=i\tau^3\gamma^{\dot{5}}\psi_m.
\]

We can compute the gauge algebra for the multiplet (\ref{mino}). It is 
the projection of the five dimensional gauge algebra:
\begin{equation}
\begin{split}
[\delta_Q(\varepsilon),\delta _Q (\eta)]  = &  
\delta_{g.c.}(i\bar{\eta}\gamma^m\varepsilon)+\delta_Q(i\bar{\varepsilon}\gamma^m\eta\psi_m)+\delta_{ch}(i\bar{\varepsilon}\gamma^m\eta 
a_m)\\ 
+ & 
\delta_{Lt}(i\bar{\eta}\gamma^m\varepsilon\widehat{\omega}_{mab}-2i\bar{\varepsilon}\gamma_{abc}\gamma^{\dot{5}}\eta 
b^c+4\bar{\varepsilon}\tau^1\gamma_{ab}\eta t^1+4\bar{\varepsilon}\tau^2\gamma_{ab}\eta 
t^2).\label{algorbi}
\end{split}
\end{equation}
$\delta_{ch}(\alpha)$ denotes a chiral transformation with parameter 
$\alpha$ and all transformation have to be understood in the four 
dimensional sense.

We are now in a position to elaborate on the matter multiplets which 
may be coupled to the supergravity multiplet (\ref{mino}). This has been 
done in great detail in \cite{Sohnius:1983xs}. We restrict ourselves to 
the presentation of formulas which are required for the remaining 
sections. A more extended discussion may be found in the original work, for 
details using our notations, \cite{diss} should be consulted.

\subsection{The chiral multiplet}\label{chm}
This multiplet has also been given in \cite{Sohnius:1983xs}. Since it 
is very well known we restrict ourselves to the transformation laws 
using our conventions. The field content of the chiral multiplet is
\[
\mathbb{A}=(A,~B,~\psi,~F,~G).
\]
It exists for arbitrary chiral weight $w$. The transformation laws are:
\begin{eqnarray*}
\delta A & = & \bar{\varepsilon}\tau^2 \psi+wB\alpha\\
 \delta B &  = & \bar{\varepsilon}\tau^2 
\gamma^{\dot{5}}\psi-wA\alpha\\
\delta \psi & = &-\frac{i}{2}\gamma^a\tau^2 \varepsilon\widehat{\cal 
D}_a A-\frac{1}{2}\varepsilon F-\frac{i}{2}\gamma^a\gamma^{\dot{5}}\tau^2  
\varepsilon \widehat{\cal D}_a B 
-\frac{1}{2}\gamma^{\dot{5}}\varepsilon G+(w-1)\gamma^{\dot{5}}\psi \alpha\\
\delta F & = & i\bar{\varepsilon}\gamma^a(\widehat{\cal 
D}_a\psi+\gamma^{\dot{5}}\psi b_a) +4 \bar{\varepsilon}(\tau^1\psi 
t^1-\tau^1\gamma^{\dot{5}}\psi t^2)\\
& + & \frac{w}{2}\bar{\varepsilon}\tau^2 ( A 
+\gamma^{\dot{5}}B)\lambda'+(2-w)G\alpha\\
\delta G & = & i\bar{\varepsilon}\gamma^{\dot{5}}\gamma^a(\widehat{\cal 
D}_a\psi+\gamma^{\dot{5}}\psi b_a)  
-4\bar{\varepsilon}\gamma^{\dot{5}}(\tau^1\psi t^1-\tau^1\gamma^{\dot{5}}\psi t^2)\\
& + & \frac{w}{2}\bar{\varepsilon}\tau^2 \gamma^{\dot{5}}( A + 
\gamma^{\dot{5}}B)\lambda'+(w-2)F\alpha.
\end{eqnarray*}
We have also indicated the transformation properties under chiral 
rotations.

For a chiral multiplet with weight $w=2$ we can write down an invariant 
$F$-term density:
\begin{equation}
[\mathbb{A}]_F = 
F+i\bar{\psi}_m\gamma^m\psi+\frac{1}{2}\bar{\psi}_m\tau^2 \gamma^{mn}(A+\gamma^{\dot{5}}B)\psi_n-12 t^2A-12 
t^1B\label{chiral}.   
\end{equation}

\subsection{The vector multiplet}\label{vec}
The field content of the super Yang-Mills multiplet is
\[
\mathbb{Y}=(~u_m,~\chi,~ D).
\]
The abelian vector multiplet can be derived straightforwardly from the 
general multiplet. We skip this multiplet, refer the interested reader 
for technical details to \cite{diss} and confine ourselves to the 
presentation of the transformation laws of the vector multiplet:
  \begin{eqnarray*}
\delta u_m & = & i\bar{\varepsilon}\gamma_m\chi\\
\delta \chi & = & -\gamma^{\dot{5}}\varepsilon D + 
\frac{1}{4}\gamma^{ab}\varepsilon \widehat{u}_{ab}\\
\delta D & = 
&\frac{i}{2}\bar{\varepsilon}\gamma^{\dot{5}}\gamma^m\widehat{\cal D}_m\chi-\frac{3i}{2}\bar{\varepsilon}\gamma^a\chi b_a,
\end{eqnarray*}
where the fields take their values in the Lie algebra of the gauge 
group. We have defined the supercovariant field strength $\widehat{u}_{mn}$ 
of $u_m$. Of course, the gauge algebra (\ref{algorbi}) is modified by 
the appearance of a field dependent gauge transformation.

An action can be derived by forming a chiral multiplet from fields of 
the Yang-Mills multiplet. One finds \cite{Sohnius:1983xs, diss}
\begin{equation}
\begin{split}
{\cal L}_{sym} = Tr~ [ & D D 
-\frac{1}{4}\widehat{u}_{ab}\widehat{u}^{ab}-3i\bar{\chi}\gamma^a\gamma^{\dot{5}}\chi b_a\\
+ & i\bar{\chi}\gamma^a\widehat{\cal D}_a\chi
- \bar{\psi}_m\gamma^m(i\gamma^{\dot{5}}\chi 
D+\frac{i}{4}\gamma^{ab}\chi\widehat{u}_{ab})\\
 + & 
\frac{1}{4}\bar{\psi}_m\tau^2\gamma^{mn}\psi_n\bar{\chi}\tau^2\chi+\frac{1}{4}\bar{\psi}_m\tau^2\gamma^{mn}\gamma^{\dot{5}}\psi_n\bar{\chi}\tau^2\gamma^{\dot{5}}\chi].\label{lqed}
\end{split}
\end{equation}
\section{Lagrangians at the fixpoints}\label{chap4}
In this section we construct lagrangians of the Randall-Sundrum type, 
i.e. theories which live on an orbifold $S^1/\mathbb{Z}_2$. In the bulk 
there is only gauged supergravity, but this can be extended easily to 
more complicated configurations using the results in 
\cite{Zucker:1999fn, diss}. On the boundaries, the original Randall-Sundrum scenario 
\cite{Randall:1999ee} requires a cosmological constant. We present some 
generalizations of this model. 
\subsection{A cosmological constant -- The Randall-Sundrum 
scenario}\label{cosm}
To start with the supersymmetrization of the Randall-Sundrum scenario 
we analyze the parity assignments of the tensor multiplet from section 
\ref{gos}. The results are shown in table \ref{table22}.
\begin{table}
\begin{center}
\begin{tabular}{|c|c|c|c|c|c|c|c|}
\hline\hline

Field & $Y^1$ & $Y^2$ & $Y^3$& $\rho$ & $N$ & $B_{mnp}$ & $B_{mn5}$\\
\hline
Parity $\cal P$ & $+1$ & $+1$ & $-1$ & $+1$ &+1 & $+1$ &$-1$\\
\hline\hline
\end{tabular}
\end{center}
\caption{Parities of the  tensor multiplet.}\label{table22}
\end{table}
Inspection of the transformation laws shows, that on the boundaries, 
they form chiral multiplets with chiral weight $w=2$. 
The precise correspondence is
\begin{equation}
\big(A,~B,~\psi,~F,~G\big)= \big(~Y^2,~Y^1,~\rho,~-2N+\widehat{\cal 
D}_{\dot{5}}Y^3, ~+2 W^{\dot{5}}+12(t^1Y^2-Y^1t^2)\big),\label{emb}
\end{equation}
on the boundary.
 
Using the result for the $F$-term density of a chiral multiplet, eq. 
(\ref{chiral}), we are then in a position to write down the desired 
theory. Before doing that, however, we have to fix the $\cal R$-symmetry. 
This is achieved by imposing (\ref{eichfixx}) which breaks the 
$SU(2)_{\cal R}$ in the bulk to a residual $U(1)$, gauged by $V_M^2$, and breaks 
the chiral $U(1)$ on the boundaries completely. The minimal multiplet 
on the boundary is then extended to an $N=1$, $D=4$ supergravity 
multiplet with $(20+20)$ components. One then uses the expressions (\ref{emb}) 
in (\ref{chiral}) to obtain
\begin{equation}
{\cal L}_{cc}=-2N + e^u  V_{\dot 5}^1 - \bar{\rho}\tau^3\psi_{\dot 
5}+i\bar{\psi}_m\gamma^m\rho+\frac{1}{2}e^u 
\bar{\psi}_a\tau^2\gamma^{ab}\psi_b - 12 e^u t^2.\label{dM}
\end{equation}
Our complete action is
\begin{equation}
S=\int d^5x~e\left( {\cal L}_{gauged} + \Lambda_1\delta(x^5){\cal 
L}_{cc}+\Lambda_2\delta(x^5-\ell){\cal L}_{cc}\right)\label{rs2}
\end{equation}
where $\Lambda_1$ and $\Lambda_2$ are real constants. We use the 
unusual convention that $\delta(x^5)$ is a scalar, not a density. If one 
eliminates the auxiliary fields, terms of the form $\delta(x^5)^2$ appear 
but drop out at the end. One finds a result similar to the one of 
Altendorfer et al. \cite{Altendorfer:2000rr},
\begin{equation}
{\cal L}=\widetilde{\cal L}_{gauged}- 
\big(\Lambda_1\delta(x^5)+\Lambda_2\delta(x^5-\ell)\big)\big( 
\sqrt{3}g'-\frac{1}{2}\bar{\psi}_a\tau^2\gamma^{ab}\psi_b\big),\label{rslag}
\end{equation}
with the on-shell lagrangian $\widetilde{\cal L}_{gauged}$ of gauged 
supergravity given in \cite{Zucker:1999fn}. $g'$ is the cosmological 
constant.
The on-shell transformation laws are
\begin{equation}
\begin{split}
\delta e_M^A = & - i\bar{\varepsilon}\gamma^A\psi_M\\
\delta \psi_M = &  {\cal D}_M \varepsilon 
+\frac{1}{4\sqrt{3}}\gamma_{MAB}\varepsilon\widehat{F}^{AB}-\frac{1}{\sqrt{3}}\gamma^N\varepsilon\widehat{F}_{MN}+\frac{ig'}{2\sqrt{3}}\gamma_M\tau^2\varepsilon\\
& + i\Lambda_1\delta_M^5\delta(x^5)\tau^2\gamma^{\dot{5}}\varepsilon 
e_5^{\dot{5}}+i\Lambda_2\delta_M^5\delta(x^5-\ell)\tau^2\gamma^{\dot{5}}\varepsilon 
e_5^{\dot{5}}\\
\delta A_M  = &  -  \frac{\sqrt{3}i}{2}\bar{\varepsilon}\psi_M.
\end{split}\label{traflaw}
\end{equation}
As a consistency check, we have verified the invariance of the 
lagrangian (\ref{rslag}) under the on-shell transformations (\ref{traflaw}).

One of the motivations for the present work where the hope to remove 
the fine tuning of the bulk and boundary cosmological constants, inherent 
to the original work of Randall and Sundrum, by supersymmetry. It 
should be clear that the method of Altendorfer et al. 
\cite{Altendorfer:2000rr} is not suited to answer the question whether this fine tuning is 
removed by supersymmetry. The alternative work \cite{Falkowski:2000er} 
implies that this fine tuning is removed by supersymmetry. This is not 
true for our result (\ref{rslag}), at least on the lagrangian level, due 
to the appearance of the parameters $\Lambda_1$ and $\Lambda_2$.

\medskip

Before proceeding, let us shortly comment on the solutions to the 
equation of motion and the Killing spinor equations. The Einstein equations 
are solved by the Randall-Sundrum metric
\begin{equation}
ds^2 = e^{-2\sigma}\eta_{mn}dx^mdx^n-(d{x^5})^2\qquad  \mbox{with} 
\qquad\sigma\equiv \frac{g'}{\sqrt{3}}|x^5|,\label{rsmetric}
\end{equation}
provided that 
\begin{equation}
\Lambda_1=-\Lambda_2=1.\label{val2}
\end{equation}
Note that we have used according to common practice, that\footnote{A 
counterexample is found as follows: Taking the derivative of 
$\theta(x^5)^2=1$ leads to $\delta(x^5)\theta(x^5)=0$ and multiplying this equation 
with $\theta(x^5)$ leads, using $\theta(x^5)^2=1$ to $\delta(x^5)=0$. 
More compactly stated: 
$\big(\theta(x^5)\theta(x^5)\big)\delta(x^5)\neq\theta(x^5)\big(\theta(x^5)\delta(x^5)\big)$. See also 
\cite{Conrad:1998ww}} $\theta(x^5)^2=1$, where $\theta$ is the standard step function.

By decomposing the spinors in two-component objects
\begin{equation}
\varepsilon^i=\binom{\varepsilon_+^i}{\varepsilon_-^i}.\label{zerl}
\end{equation}
and using the representation of gamma matrices given in 
\cite{Mirabelli:1998aj}, one may try to solve the Killing spinor equations 
$\delta\psi_M=0$ using the Ansatz of Altendorfer et al. \cite{Altendorfer:2000rr}
\begin{equation}
\varepsilon_+^1=e^{-\frac{1}{2}\sigma}\eta,\qquad 
\varepsilon_+^2=-i\theta(x^5)e^{-\frac{1}{2}\sigma}\eta,\label{damned}
\end{equation}
with a constant spinor $\eta$. However, in this case one finds the 
constraint 
\begin{equation}
\Lambda_1 = -\Lambda_2 =2.\label{val1}
\end{equation}
Note that (\ref{val1}) is in contradiction with (\ref{val2}) and thus 
the spinors (\ref{damned}) represent no Killing spinors of the 
Randall-Sundrum solution.\footnote{In equation (13) of 
\cite{Altendorfer:2000rr}, the transformation law for the gravitino $\psi_m^+$ contains a term 
$\sim i\Lambda e_m{}^a\sigma_a\bar{\eta}_+$ which is missing in eq. 
(20). If this term is included, eqs. (21) and (22) have to be changed. 
After performing these corrections, one recovers the $\Lambda_1 = 
-\Lambda_2 =2$ case and our calculation is in agreement with 
\cite{Altendorfer:2000rr}. We would like to thank Jan Conrad for extended discussions on 
this point.} However, since the Randall-Sundrum metric (\ref{rsmetric}) 
is a solution to the Einstein equations we have to expect that there 
exist consistent solutions to the Killing spinor equations. Otherwise we 
would have constructed a theory where the geometry of spacetime breaks 
supersymmetry spontaneously.

\subsection{Matter at the fixpoints}\label{matt}

Our next models contain as bulk theory gauged supergravity and 
cosmological constants on the boundaries. In addition, we allow for chiral 
multiplets or super Yang-Mills multiplets on the boundaries. Let us start 
with the chiral multiplet.

For notational convenience we consider only one brane located at 
$x^5=0$. It is completely trivial to add the second one.
\medskip

The systematic construction of a kinetic term for the chiral multiplet 
as it is needed here requires the general multiplet which we do not 
discuss. For technical details we refer to \cite{Sohnius:1983xs, diss}. 
The bosonic part of the lagrangian for a chiral multiplet with chiral 
weight $w$ is
\begin{equation}
\begin{split}
{\cal L}_{kin} = & -{\cal D}_a B {\cal D}^a B - {\cal D}_a A {\cal D}^a 
A 
-8t^2 GB- F^2-G^2\\
+ & 8t^2 FA+8t^1FB + 8t^1GA-4{\cal D}_a  BA b^a+4{\cal D}_a  AB b^a\\
+ & 
(A^2+B^2)(12wS-\frac{w}{4}R_{ab}{}^{ab}+48(w-1)((t^1)^2+(t^2)^2)-6wb_ab^a-8S).
\end{split}\label{kinch}
\end{equation}
Note the appearance of the curvature scalar. After inserting the 
definitions (\ref{wichtig1}) and fixing the gauge via (\ref{eichfixx}), we 
take our action to be
\[
S=\int d^5x~e\left({\cal L}_{gauged}+\Lambda_1\delta(x^5){\cal 
L}_{cc}+\tilde{\Lambda}_1\delta(x^5){\cal L}_{kin}\right),
\]
with $\tilde{\Lambda}_1$ a constant.
After gauge fixing, the final step would be the elimination of the 
auxiliary fields. This turns out to be a remarkably complicated exercise 
which we have not completed.

\medskip

The other possibility is to consider a super Yang-Mills multiplet on 
the boundary. This is conceptually similar to the model of Ho\v{r}ava and 
Witten \cite{Horava:1996ma, Horava:1996qa}.

Our model is the following. In the bulk, we have gauged supergravity. 
On the boundary there is a cosmological constant plus a super Yang-Mills 
multiplet. Our action is
\[
S=\int d^5x~e\left({\cal L}_{gauged}+\Lambda_1\delta(x^5){\cal 
L}_{cc}+\tilde{\Lambda}_1\delta(x^5){\cal L}_{sym}\right)
\]
with $\tilde{\Lambda}_1$ a constant and ${\cal L}_{sym}$ is given in 
(\ref{lqed}). We next use the definitions (\ref{wichtig1}) and fix the 
gauge using (\ref{eichfixx}). 
In this case it is easy to eliminate the auxiliary fields from the 
lagrangian. The result is
\begin{equation}
\begin{split}
{\cal L} & = \widetilde{\cal L}_{gauged}-\Lambda_1 
\delta(x^5)(\sqrt{3}g'-\frac{1}{2}\bar{\psi}_a\tau^2\gamma^{ab}\psi_b)\\
+ & 
\tilde{\Lambda}_1\delta(x^5)\big(-\frac{1}{4}\widehat{u}_{ab}\widehat{u}^{ab}-\frac{i\sqrt{3}}{4}\bar{\chi}\gamma^a\gamma^{\dot{5}}\chi\widehat{F}_{a\dot{5}}+i\bar{\chi}\gamma^m\widehat{D}_m\chi-\frac{i}{2}\bar{\chi}\gamma^{mab}\psi_m\widehat{u}_{ab}\\
+ & \frac{1}{4}\bar{\psi}_m\tau^2\gamma^{mn}\psi_n\bar{\chi}\tau^2\chi 
+ 
\frac{1}{4}\bar{\psi}_m\tau^2\gamma^{mn}\gamma^{\dot{5}}\psi_n\bar{\chi}\tau^2\gamma^{\dot{5}}\chi\big).
\end{split}\label{sqedrs}
\end{equation}

\section{Conclusions and Outlook}\label{conc}
Using off-shell supergravity, we have presented a technique which 
allows the systematic construction of models of the Randall-Sundrum type. We 
assign consistently $\mathbb{Z}_2$ transformation laws to the fields 
belonging to the supergravity multiplet in the bulk. The fields with 
positive parity form a non-minimal $N=1$, $D=4$ supergravity multiplet at 
the fixpoints. Using this multiplet, we develop parts of an $N=1$ tensor 
calculus located at the branes. We apply this calculus to the explicit 
construction of lagrangians of the Randall-Sundrum type. 

The configurations presented here are easily extended to more 
complicated ones, using the results in \cite{Zucker:2000ej, Zucker:1999fn, 
diss}. That is, one could introduce additional matter to the bulk which 
would then extend the possible couplings to matter at the fixpoints of the 
orbifold.

Of particular interest would be the study of supersymmetry breaking, 
since the brane world scenario represents a geometrization of hidden 
sector supergravity models (for a review see \cite{Nilles:1984ge}). 

First steps in this direction were undertaken in 
\cite{Mirabelli:1998aj} and in \cite{gabi}. These authors considered rigidly supersymmetric 
theories on $S^1/\mathbb{Z}_2$. On one brane a supersymmetry breaking 
$D$- or $F$-term was placed, respectively, and the transmission of the 
supersymmetry breaking to the other wall studied. Using the techniques 
developed in the present work, it should be quite simple to extend these 
investigations to local supersymmetry. In this context we should also 
mention \cite{Falkowski:2000er}, where gaugino condensation as 
supersymmetry breaking mechanism has been considered.

\subparagraph{Acknowledgments:}
We would like to thank Jonathan Bagger, Jan Conrad, Stefan F\"orste, 
Zygmunt Lalak and Hans-Peter Nilles for helpful discussions.
This work was partially supported by the European Commission programs 
ERBFMRX-CT96-0045 and CT96-0090.

\end{document}